\pdfoutput=1

\documentclass{jors}

\begin{document}

{\bf Software paper for submission to the Journal of Open Research Software} \\

To complete this template, please replace the blue text with your own. The
paper has three main sections: (1) Overview; (2) Availability; (3) Reuse
potential.

Please submit the completed paper to: editor.jors@ubiquitypress.com

\rule{\textwidth}{1pt}

\section*{(1) Overview}

\vspace{0.5cm}

\section*{Title}


FluidFFT: common API (C++ and Python) for Fast Fourier Transform HPC libraries

\section*{Paper Authors}


1. MOHANAN Ashwin Vishnu$^a$\\
2. BONAMY Cyrille$^b$\\
3. AUGIER Pierre$^b$\\

\smallskip

$^a$ Linn\'e Flow Centre, Department of Mechanics, KTH, 10044 Stockholm, Sweden.
$^b$ Univ. Grenoble Alpes, CNRS, Grenoble INP\footnote{Institute of Engineering
Univ. Grenoble Alpes}, LEGI, 38000 Grenoble, France.\\

\section*{Paper Author Roles and Affiliations}

1. Ph.D. student, Linn\'e Flow Centre, KTH Royal Institute of Technology,
Sweden; \\
2. Research Engineer, LEGI, Universit\'e Grenoble Alpes, CNRS, France; \\
3. Researcher, LEGI, Universit\'e Grenoble Alpes, CNRS, France

\section*{Abstract}


The Python package \fluidpack{fft} provides a common Python API for performing
Fast Fourier Transforms (FFT) in sequential, in parallel and on GPU with different
FFT libraries (FFTW, P3DFFT, PFFT, cuFFT). \fluidpack{fft} is a comprehensive FFT
framework which allows Python users to easily and efficiently perform FFT and the
associated tasks, such as as computing linear operators and energy spectra.
We describe the architecture of the package composed of C++ and Cython FFT
classes, Python ``operator'' classes and Pythran functions.
The package supplies utilities to easily test itself and benchmark the different
FFT solutions for a particular case and on a particular machine.
We present a performance scaling analysis on three different computing clusters
and a microbenchmark showing that \fluidpack{fft} is an interesting solution to
write efficient Python applications using FFT.

\section*{Keywords}


Free and open-source library; Python; Fast Fourier Transform; Distributed; MPI;
GPU; High performance computing%

\section*{Introduction}


Fast Fourier Transform (FFT) is a class of algorithms used to calculate the
discrete Fourier transform, which traces back its origin to the groundbreaking
work by \citet{cooley_tukey}.
Ever since then, FFT as a computational tool has been applied in multiple
facets of science and technology, including digital signal processing, image
compression, spectroscopy, numerical simulations and scientific computing in
general. There are many good libraries to perform FFT, in particular the
\emph{de-facto} standard \libpack{FFTW} \citep{frigo2005design}.\@ A challenge
is to efficiently scale FFT on clusters with the memory distributed over a
large number of cores using Message Passing Interface (MPI). This is imperative
to solve big problems faster and when the arrays do not fit in the memory of
single computational node.
A problem is that for one-dimensional FFT, all the data have to be located in the
memory of the process that perform the FFT, so a lot of communications between
processes are needed for 2D and 3D FFT.

There are two strategies to distribute an array in the memory, the 1D (or
\emph{slab}) decomposition and the 2D (or \emph{pencil}) decomposition. The 1D
decomposition is more efficient when only few processes are used but suffers
from an important limitation in terms of number of MPI processes that can be
used. Utilizing 2D decomposition overcomes this limitation.

Some of the well-known libraries are written in C, C++ and Fortran. The classical
\libpack{FFTW} library supports MPI using 1D decomposition and hybrid parallelism
using MPI and OpenMP.  Other libraries, now implement the 2D decomposition:
\libpack{PFFT} \citep{pippig_pfft2013}, \libpack{P3DFFT}
\citep{pekurovsky2012p3dfft}, \libpack{2decomp\&FFT} and so on. These libraries
rely on MPI for the communications between processes, are optimized for
supercomputers and scales well to hundreds of thousands of cores. However, since
there is no common API, it is not simple to write applications that are able to
use these libraries and to compare their performances. As a result, developers are
met with a hard decision, which is to choose a library before the code is
implemented.

Apart from CPU-based parallelism, General Purpose computing on Graphical
Processing Units (GPGPU) is also gaining traction in scientific computing.
Scalable libraries written for GPGPU such as OpenCL and CUDA have emerged, with
their own FFT implementations, namely \libpack{clFFT} and \libpack{cuFFT}
respectively.

Python can easily link these libraries through compiled extensions. For a Python
developer, the following packages leverage this approach to perform FFT:

\begin{outline}
  \1 sequential FFT, using:
    \2 \pack{numpy.fft} and \pack{scipy.fftpack} which are essentially
    C and Fortran extensions for \libpack{FFTPACK} library.
    \2 \pack{pyFFTW} which wraps \libpack{FFTW} library and provides interfaces similar to
    the \pack{numpy.fft} and \pack{scipy.fftpack} implementations.
    \2 \pack{mkl\_fft}, which wraps Intel's \libpack{MKL} library and exposes python
    interfaces to act as drop-in replacements for \pack{numpy.fft} and
    \pack{scipy.fftpack}.
  \1 FFT with MPI, using:
    \2 \pack{mpiFFT4py} and \pack{mpi4py-fft} built on top of \pack{pyFFTW} and
    \pack{numpy.fft}.
    \2 \pack{pfft-python} which provides extensions for
    PFFT library.
  \1 FFT with GPGPU, using:
    \2 \pack{Reikna}, a pure python package which depends on \pack{PyCUDA}
    and \pack{PyOpenCL}
    \2 \pack{pytorch-fft}: provides C extensions for cuFFT, meant to work with
    PyTorch, a tensor library similar to NumPy.
\end{outline}

Although these Python packages are under active development, they suffer from
certain drawbacks:

\begin{itemize}
  \item No effort so far to consolidate all possible FFT libraries, both
  sequential, MPI and GPGPU based under a single package with similar syntax.

  \item Quite complicated even for the simplest use case scenarios. To
  understand how to use them, a novice user has to, at least, read the
  \libpack{FFTW} documentation.

  \item No benchmarks between libraries and between the Python
  solutions and solutions based only on a compiled language (as C, C++ or
  Fortran).

  \item Provides just the FFT and inverse FFT functions, no associated
  mathematical operators.

\end{itemize}

The Python package \fluidpack{fft} fills this gap by providing C++ classes and
their Python wrapper classes for performing simple and common tasks with different
FFT libraries. It has been written to make things easy while being as efficient as
possible. It provides:

\begin{itemize}
\item tests,

\item documentation and tutorials,

\item benchmarks,

\item operators for simple tasks (for example, compute the energy or the
gradient of a field).

\end{itemize}

In the present article, we shall start by describing the implementation of
\fluidpack{fft} including its design aspects and the code organization. Thereafter,
we shall compare the performance of different classes in \fluidpack{fft} in
three computing clusters, and also describe, using microbenchmarks, how a Python
function can be optimized to be as fast as a Fortran implementation. Finally,
we show how we test and maintain the quality of the code base through
continuous integration and mention some possible applications of
\fluidpack{fft}.

\section*{Implementation and architecture}

The two major design goals of \fluidpack{fft} are:
\begin{itemize}
 \item to support multiple FFT libraries under the same umbrella and expose the
 interface for both C++ and Python code development.
 \item to keep the design of the interfaces as human-centric and easy to use as
 possible, without sacrificing performance.
\end{itemize}

Both C++ and Python APIs provided by \fluidpack{fft} currently support linking
with \libpack{FFTW} (with and without MPI and OpenMP support enabled),
\libpack{MKL}, \libpack{PFFT}, \libpack{P3DFFT}, \libpack{cuFFT} libraries. The
classes in \fluidpack{fft} offers API for performing
double-precision\footnote{Most C++ classes also support single-precision.}
computation with real-to-complex FFT, complex-to-real inverse FFT, and additional
helper functions.

\subsection*{C++ API}

\begin{figure}[htp]
  \centering
  \includegraphics[width=\linewidth]{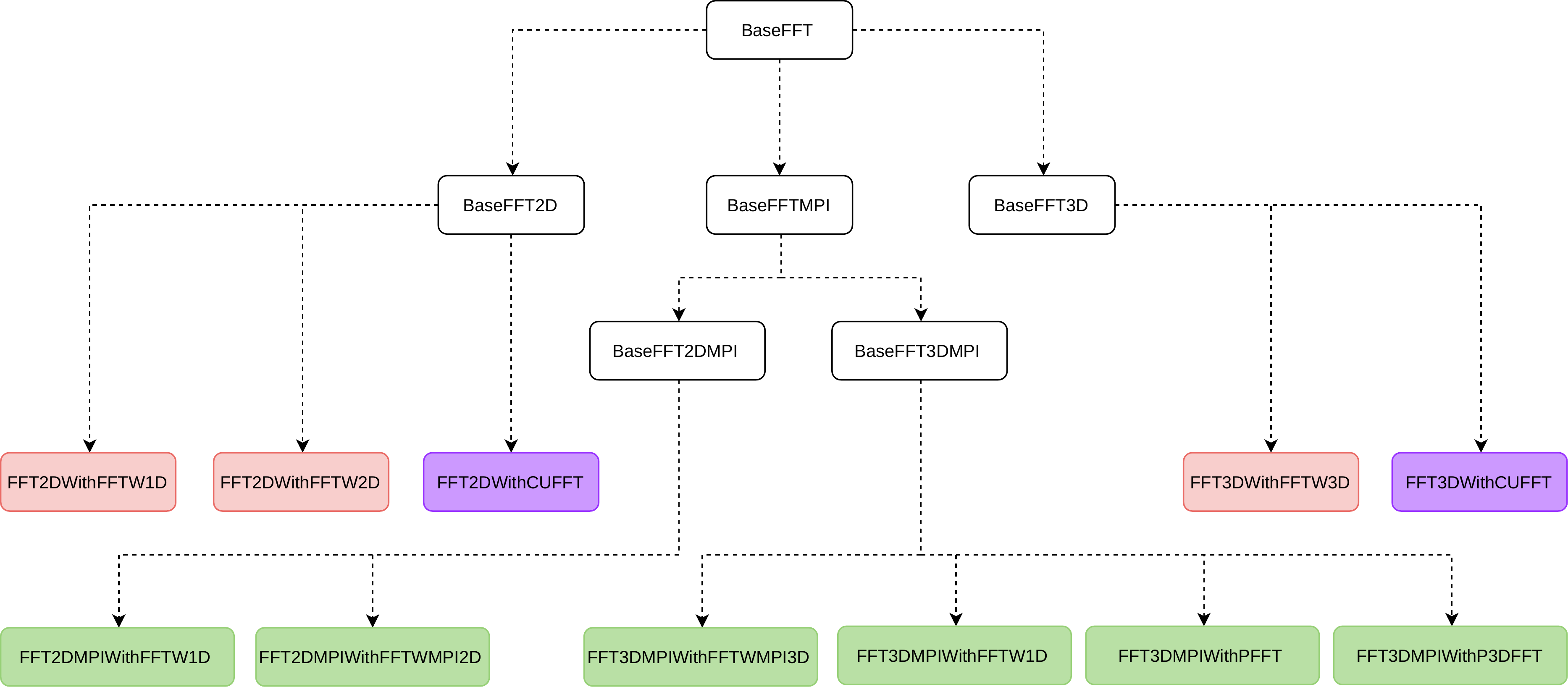}
  \caption{Class hierarchy demonstrating object-oriented approach. The
    sequential classes are shown in red, the CUDA-based classes in magenta and
    the MPI-based classes in green. The arrows represent inheritance from
    parent to child class.
  }\label{fig:classes}
\end{figure}

The C++ API is implemented as a hierarchy of classes as shown in
Fig.~\ref{fig:classes}.
The naming convention used for the classes (\codeinline{<Type of FFT>With<Name of
Library>}) is a cue for how these are functioning internally.
By utilizing inheritance, the classes share the same function names and syntax
defined in the \emph{base} classes, shown in white boxes in
Fig.~\ref{fig:classes}. Some examples of such functions are:

\begin{itemize}
  \item \codeinline{alloc\_array\_X}: Allocates array to store a physical array
    with real datatype for the current process.
  \item \codeinline{alloc\_array\_K}: Allocates array to store a spectral array
    with complex datatype  for the current process.
  \item \codeinline{init\_array\_X\_random}: Allocates and initializes a physical
    array with random values.
  \item \codeinline{test}: Run tests for a class by comparing mean and mean energy
    values in an array before and after a set of \codeinline{fft} and
    \codeinline{ifft} calls.
  \item \codeinline{bench}: Benchmark the \codeinline{fft} and
    \codeinline{ifft} methods for certain number of iterations.
\end{itemize}

Remaining methods which are specific to a library are defined in the
corresponding child classes, depicted in coloured boxes in
Fig.~\ref{fig:classes}, for example:

\begin{itemize}
  \item \codeinline{are\_parameters\_bad}: Verifies whether the global array
    shape can be decomposed with the number of MPI processes available or not.
    If the parameters are compatible, the method returns \codeinline{false}.
    This method is called prior to initializing the class.
  \item \codeinline{fft} and \codeinline{ifft}: Forward and inverse FFT
    methods.
\end{itemize}

Let us illustrate with a trivial example, wherein we initialize the FFT with a
random physical array, and perform a set of \codeinline{fft} and \codeinline{ifft}
operations.
\begin{minted}[fontsize=\footnotesize]{cpp}
#include <iostream>
using namespace std;

#include <fft3dmpi_with_fftwmpi3d.h>
// #include <fft3dmpi_with_p3dfft.h>

int main(int argc, char **argv) {
  int N0 = N1 = N2 = 32;
  // MPI-related
  int nb_procs = 4;
  MPI_Init(&argc, &argv);
  MPI_Comm_size(MPI_COMM_WORLD, &(nb_procs));

  myreal* array_X;
  mycomplex* array_K;

  FFT3DMPIWithFFTWMPI3D o(N0, N1, N2);
  // FFT3DMPIWithP3DFFT o(N0, N1, N2);

  o.init_array_X_random(array_X);
  o.alloc_array_K(array_K);
  o.fft(array_X, array_K);
  o.ifft(array_K, array_X);
  MPI_Finalize();
  return 0;
}
\end{minted}

As suggested through comments above, in order to switch the FFT library, the
user only needs to change the header file and the class name. An added
advantage is that, the user does not need to bother about the domain
decomposition while declaring and allocating the arrays. A few more helper
functions are available with the FFT classes, such as functions to compute the
mean value and energies in the array. These are demonstrated with examples in
the documentation.\footnote{%
\url{https://fluidfft.readthedocs.io/en/latest/examples/cpp.html}.}
Detailed information regarding the C++ classes and its member functions are
also included in the online documentation\footnote{%
\url{https://fluidfft.readthedocs.io/en/latest/doxygen/index.html}.}.

\subsection*{Python API} Similar to other packages in the FluidDyn project,
\fluidpack{fft} also uses an object-oriented approach, providing FFT classes.
This is in contrast with the approach adopted by \pack{numpy.fft} and \pack{%
scipy.fftpack} which provides functions instead, with which the user has to
figure out the procedure to design the input values and to use the return
values, from the documentation.
In \fluidpack{fft}, the Python API wraps all the functionalities of its C++
counterpart and offers a more richer experience through an accompanying
operator class.

As a short example, let us try to calculate the gradient of a plane sine-wave
using spectral methods, mathematically described as follows:

\begin{align*}
  u(x,y) &=
    \sin(x + y) &\forall x,y \in \left[0, L \right] \\
  \hat u(k_x,k_y) &=
    \frac{1}{L^2}
    \int_0^{L}\int_0^{L}
    u(x,y) \exp(ik_x x + ik_y y) dx dy \\
  \nabla u(x,y) &=
    \sum_{k_x} \sum_{k_y}
    i\mathbf{k}
    \hat u(k_x,k_y) \exp(-ik_x x - ik_y y)
\end{align*}
where $k_x$, $k_y$ represent the wavenumber corresponding to x- and y-directions,
and $\mathbf{k}$ is the wavenumber vector.

The equivalent pseudo-spectral implementation in \fluidpack{fft} is as follows:
\begin{minted}[fontsize=\footnotesize]{python}
  from fluidfft.fft2d.operators import OperatorsPseudoSpectral2D, pi
  from numpy import sin

  nx = ny = 100
  lx = ly = 2 * pi

  oper = OperatorsPseudoSpectral2D(nx, ny, lx, ly, fft="fft2d.with_fftw2d")

  u = sin(oper.XX + oper.YY)
  u_fft = oper.fft(u)
  px_u_fft, py_u_fft = oper.gradfft_from_fft(u_fft)
  px_u = oper.ifft(px_u_fft)
  py_u = oper.ifft(py_u_fft)
  grad_u = (px_u, py_u)
\end{minted}

A parallelized version of the code above will work out of the box, simply by
replacing the FFT class with an MPI-based FFT class, for instance
\codeinline{fft2d.with\_fftwmpi2d}. One can also let \fluidpack{fft} automatically
choose an appropriate FFT class by instantiating the operator class with
\codeinline{fft=None} or \codeinline{fft="default"}. Even if one finds the methods
in the operator class to be lacking, one can inherit the class and easily create a
new method, for instance using the wavenumber arrays, \codeinline{oper.KX} and
\codeinline{oper.KY}.  Arguably, a similar implementation with other available
packages would require the know-how on how FFT arrays are allocated in the memory,
normalized, decomposed in parallel and so on.
Moreover, the FFT and the operator classes contain objects describing the shapes
of the real and complex arrays and how the data is shared between processes.
A more detailed introduction on how
to use \fluidpack{fft} and available functions can be found in the
tutorials\footnote{%
\url{https://fluidfft.readthedocs.io/en/latest/tutorials.html}.}.

Thus, we have demonstrated how, by using \fluidpack{fft}, a developer can
easily switch between FFT libraries.
Let us now turn our attention to how the code is organized. We shall also describe
how the source code is built, and linked with the supported libraries.

\subsection*{Code organization}
\begin{figure}[htp]
  \centering
  \includegraphics[width=0.96\linewidth]{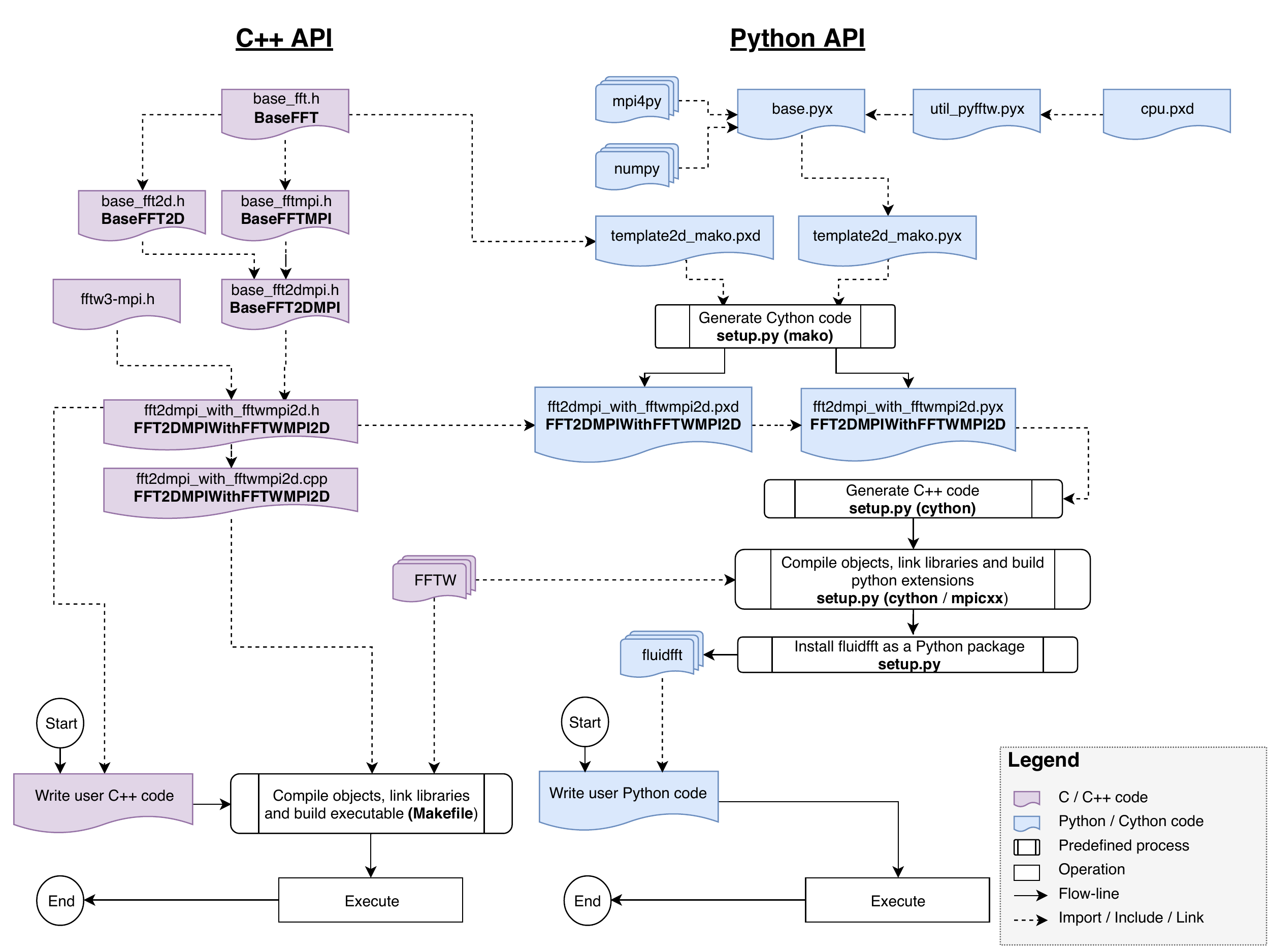}
  \caption{Flowchart illustrating how the C++ and Python API are built and used
  for one particular class, viz. \codeinline{FFT2DMPIWithFFTWMPI2D}. The dotted
  arrows in C++ part stand for include statements, demonstrating the class
  hierarchy and in the Python part indicate how different codes are imported. On
  the bottom, a smaller flowchart demonstrates how to use the API by writing user
  code.  }\label{fig:build_use}
\end{figure}

The internal architecture of \fluidpack{fft} can be visualized as layers.  Through
Fig.~\ref{fig:build_use}, we can see how these layers are linked together forming
the API for C++ and Python development. For simplicity, only one FFT class is
depicted in the figure, namely \codeinline{FFT2DMPIWithFFTWMPI2D}, which wraps
\libpack{FFTW}'s parallelized 2D FFT implementation. The C++ API is accessed by
importing the appropriate header file and building the user code with a Makefile,
an example of which is available \href{%
https://fluidfft.readthedocs.io/en/latest/examples/cpp.html}{%
in the documentation}.

The Python API is built automatically when \fluidpack{fft} is
installed\footnote{%
\href{https://fluidfft.readthedocs.io/en/latest/install.html}{Detailed steps
for installation} are provided in the documentation.}.
It first generates the Cython source code as a pair of \codeinline{.pyx} and
\codeinline{.pxd} files containing a class wrapping its C++
counterpart\footnote{Uses an approach similar to guidelines \href{%
    https://cython.readthedocs.io/en/latest/src/userguide/wrapping_CPlusPlus.html}{%
``Using C++ in Cython''} in the Cython documentation.}.
The Cython files are produced from template files (specialized for the 2D and
3D cases) using the template library \mako.
Thereafter, \pack{Cython} \citep{behnel_cython2011} generates C++ code with
necessary Python bindings, which are then built in the form of extensions or
dynamic libraries importable in Python code. All the built extensions are then
installed as a Python package \fluidpack{fft}.

A helper function \codeinline{fluidfft.import\_fft\_class} is provided with the
package to simply import the FFT class. However, it is more convenient and
recommended to use an operator class, as described in the example for Python
API.\@ Although the operator classes can function as pure Python code, some of
its critical methods can be compiled, if \pack{Pythran}
\citep{guelton2018pythran} is available during installation of
\fluidpack{fft}. We will show towards the end of this section that by using
\pack{Pythran}, we reach the performance of the equivalent Fortran code.

To summarize, \fluidpack{fft} consists of the following layers:
\begin{itemize}

\item One C++ class per method derived from a hierarchy of C++ classes as shown
in Fig.~\ref{fig:classes}.

\item \pack{Cython} wrappers of the C++ classes with their unit test cases.

\item Python operators classes (2D and 3D) to write code independently of the
library used for the computation of the FFT and with some mathematical helper
methods. These classes are accompanied by unit test cases.

\item \pack{Pythran} functions to speedup critical methods in the Python
operators classes.

\end{itemize}

Command-line utilities (\codeinline{fluidfft-bench} and
\codeinline{fluidfft-bench-analysis}) are also provided with the \fluidpack{fft}
installation to run benchmarks and plot the results. In the next subsection, we
shall look at some results by making use of these utilities on three computing
clusters.

\subsection*{Performance}

\subsubsection*{Scalability tests using \codeinline{fluidfft-bench}}


Scalability of \fluidpack{fft} is measured in the form of strong scaling speedup,
defined in the present context as:
\begin{equation*}
S_\alpha(n_p) = \frac
{[\mathrm{Time\ elapsed\ for\ } N \mathrm{\ iterations\ with\ }n_{p,\min}\mathrm{\ processes}]_{\mathrm{fastest}}
\times n_{p,\min}}
{[\mathrm{Time\ elapsed\ for\ } N \mathrm{\ iterations\ with\ } n_p \mathrm{\
processes}]_\alpha}
\label{eq:speedup}
\end{equation*}

where $n_{p,\min}$ is the minimum number of processes employed for a specific
array size and hardware. The subscripts, $\alpha$ denotes the FFT class used and
``fastest'' corresponds to the fastest result among various FFT classes.

To compute strong scaling the utility \codeinline{fluidfft-bench} is launched
as scheduled jobs on HPC clusters, ensuring no interference from background
processes. No hyperthreading was used.
We have used $N=20$ iterations for each run, with which we obtain sufficiently
repeatable results.
For a particular choice of array size, every FFT class available are
benchmarked for the two tasks, forward and inverse FFT. Three different function
variants are compared (see the legend in subsequent figures):

\begin{itemize}

\item \codeinline{fft\_cpp}, \codeinline{ifft\_cpp} (continuous lines): benchmark
of the C++ function from the C++ code. An array is passed as an argument to store
the result. No memory allocation is performed inside these functions.

\item \codeinline{fft\_as\_arg}, \codeinline{ifft\_as\_arg} (dashed lines):
benchmark of a Python method from Python. Similar to the C++ code, the second
argument of this method is an array to contain the result of the transform, so no
memory allocation is needed.

\item \codeinline{fft\_return}, \codeinline{ifft\_return} (dotted lines):
benchmark of a Python method from Python. No array is provided to the function to
contain the result, and therefore a numpy array is created and then returned by
the function.

\end{itemize}

On big HPC clusters, we have only focussed on 3D array transforms as benchmark
problems, since these are notoriously expensive to compute and require massive
parallelization.  The physical arrays used in all four 3D MPI based FFT classes
are identical in structure.  However, there are subtle differences, in terms of
how the domain decomposition and the allocation of the transformed array in the
memory are handled\footnote{Detailed discussion on \href{%
https://fluidfft.readthedocs.io/en/latest/ipynb/executed/tuto_fft3d_mpi_domain_decomp.html}{%
``FFT 3D parallel (MPI): Domain decomposition''} tutorial}.

Hereafter, for the sake of brevity, the FFT classes will be named in terms of the
associated library (For example, the class \codeinline{FFT3DMPIWithFFTW1D} is
named \codeinline{fftw1d}).  Let us go through the results\footnote{Saved at
\url{%
https://bitbucket.org/fluiddyn/fluidfft-bench-results}} plotted using
\codeinline{fluidfft-bench-analysis}.

\paragraph{Benchmarks on Occigen}

\href{https://www.top500.org/system/178465}{Occigen} is a GENCI-CINES HPC
cluster which uses Intel Xeon CPU E5--2690 v3 (2.6 GHz) processors with 24 cores
per node. The installation was performed using Intel C++ 17.2 compiler, Python
3.6.5, and OpenMPI 2.0.2.

\begin{figure}[htp!]
\centering
\includegraphics[width=\linewidth]{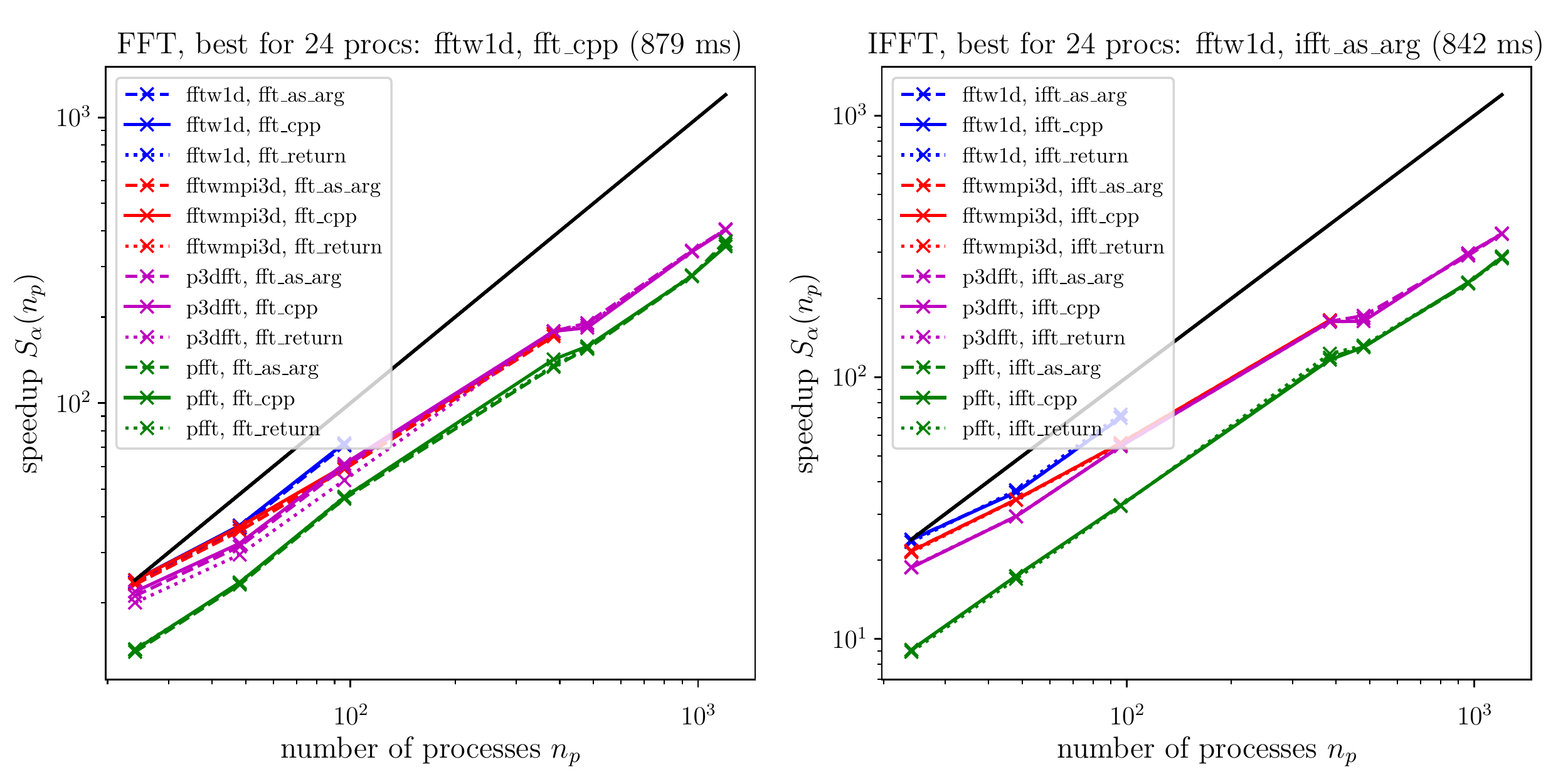}
\caption{Speedup computed from the median of the elapsed times for 3D fft
(384$\times$1152$\times$1152, left: fft and right: ifft) on Occigen.}%
\label{fig:occigen384x1152x1152}
\end{figure}

Fig.~\ref{fig:occigen384x1152x1152} demonstrates the strong scaling performance
of a cuboid array sized $384\times1152\times1152$. This case is particularly
interesting since for FFT classes implementing 1D domain decomposition
(\codeinline{fftw1d} and \codeinline{fftwmpi3d}), the processes are spread on
the first index for the physical input array. This restriction is as a result
of some \libpack{FFTW} library internals and design choices adopted in
\fluidpack{fft}. This limits \codeinline{fftw1d} to 192 cores and
\codeinline{fftwmpi3d} to 384 cores. The latter can utilize more cores since it
is capable of working with empty arrays, while sharing some of the
computational load.
%
The fastest methods for relatively
low and high number of processes are \codeinline{fftw1d} and
\codeinline{p3dfft} respectively for the present case.

The benchmark is not sufficiently accurate to measure the cost of calling the
functions from Python (difference between continuous and dashed lines,
i.e. between pure C++ and the \codeinline{as\_arg} Python method) and even the
creation of the numpy array (difference between the dashed and the dotted line,
i.e. between the \codeinline{as\_arg} and the \codeinline{return} Python
methods).

\begin{figure}[htp!]
\centering
\includegraphics[width=\linewidth]{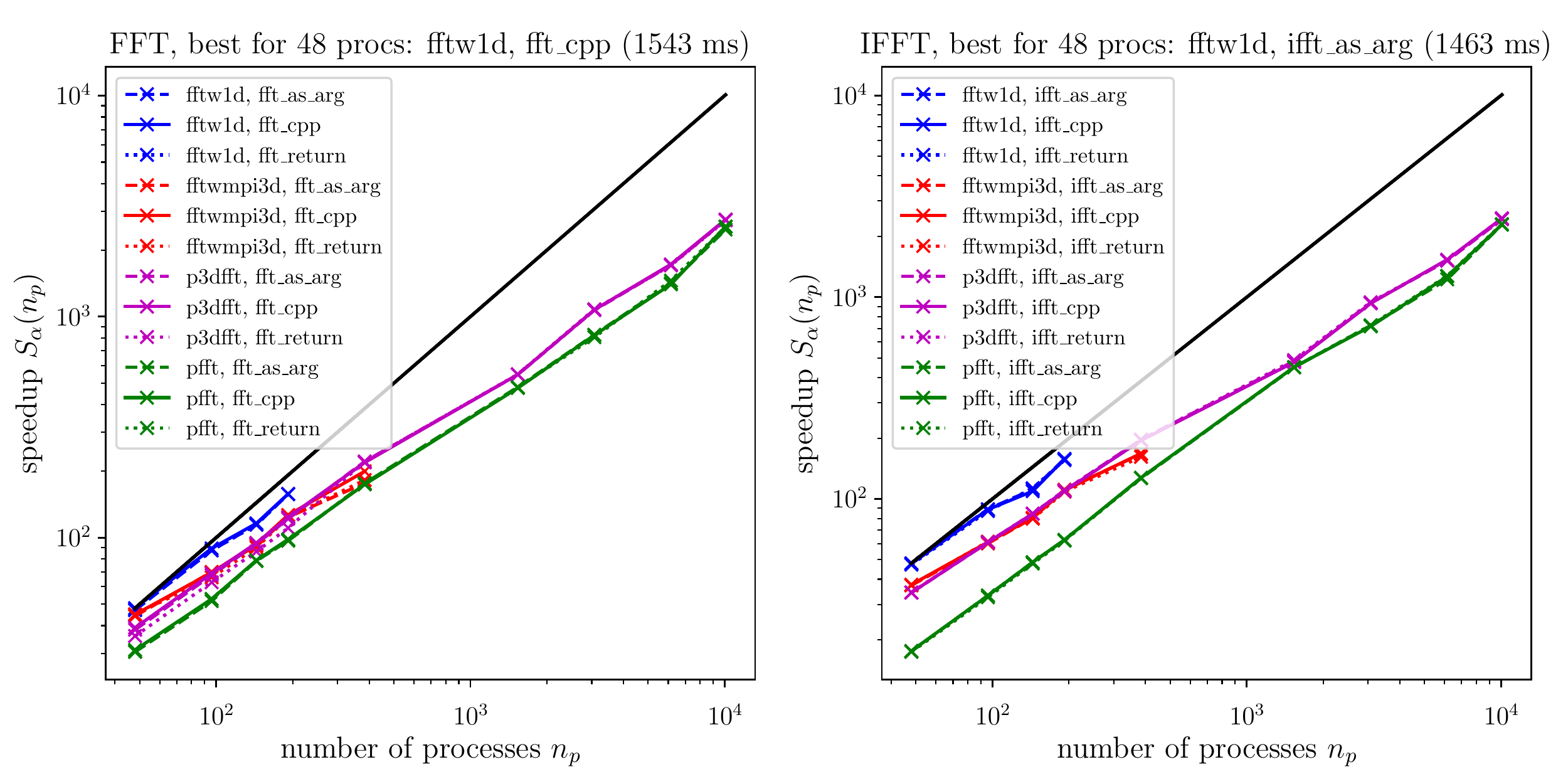}
\caption{Speedup computed from the median of the elapsed times for 3D fft
(1152$\times$1152$\times$1152, left: fft and right: ifft) on Occigen.}
\label{fig:occigen1152x1152x1152}
\end{figure}

Fig.~\ref{fig:occigen1152x1152x1152} demonstrates the strong scaling
performance of a cubical array sized $1152\times1152\times1152$. For this
resolution as well, \codeinline{fftw1d} is the fastest method when using only
few cores and it can not be used for more that 192 cores. The faster library
when using more cores is also \codeinline{p3dfft}. This also shows that
\fluidpack{fft} can effectively scale for over 10,000 cores with a significant
increase in speedup.

\paragraph{Benchmarks on Beskow}

\href{ https://www.pdc.kth.se/hpc-services/computing-systems}{Beskow} is a Cray
machine maintained by SNIC at PDC, Stockholm. It runs on Intel(R) Xeon(R) CPU
E5-2695 v4 (2.1 GHz) processors with 36 cores per node. The installation was
done using Intel C++ 18 compiler, Python 3.6.5 and CRAY-MPICH 7.0.4.

\begin{figure}[htp!]
\centering
\includegraphics[width=\linewidth]{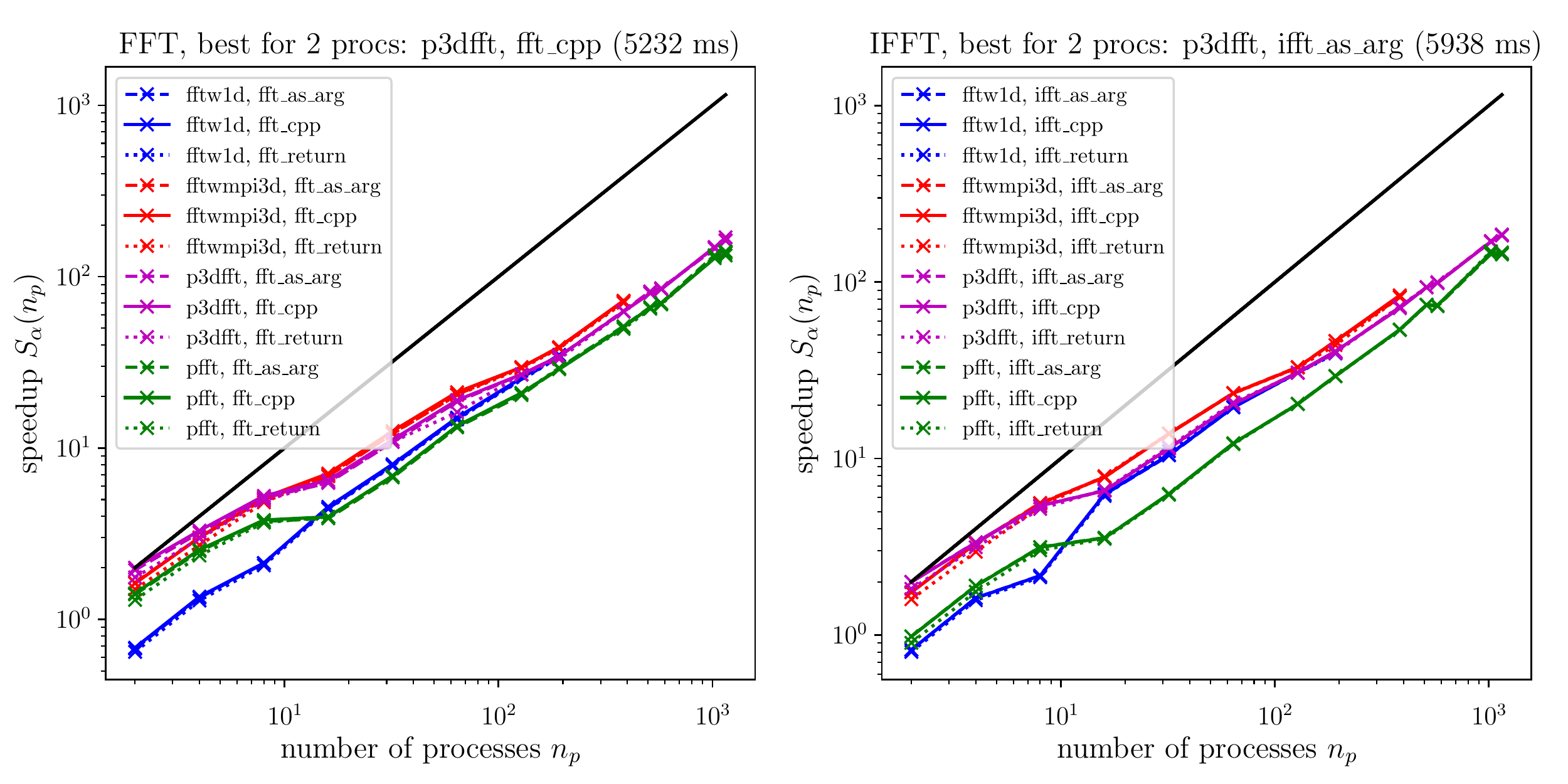}
\caption{Speedup computed from the median of the elapsed times for 3D fft
(384$\times$1152$\times$1152, left: fft and right: ifft) on Beskow.}
\label{fig:beskow384x1152x1152}
\end{figure}

In Fig.~\ref{fig:beskow384x1152x1152}, the strong scaling results of the cuboid
array can be observed. In this set of results we have also included intra-node
scaling, wherein there is no latency introduced due to typically slower
node-to-node communication. The fastest library for very low (below 16) and
very high (above 384) number of processes in this configuration is
\codeinline{p3dfft}. For moderately high number of processes (16 and above) the
fastest library is \codeinline{fftwmpi3d}. Here too, we notice that
\codeinline{fftw1d} is limited to 192 cores and \codeinline{fftwmpi3d} to 384
cores, for reasons mentioned earlier.

A striking difference when compared with Fig.~\ref{fig:occigen384x1152x1152} is
that \codeinline{fftw1d} is not the fastest of the four classes in this machine.
One can only speculate that this could be a consequence of the differences in MPI
library and hardware which has been employed. This also emphasises the need to
perform benchmarks when using an entirely new configuration.

\begin{figure}[htp!]
\centering
\includegraphics[width=\linewidth]{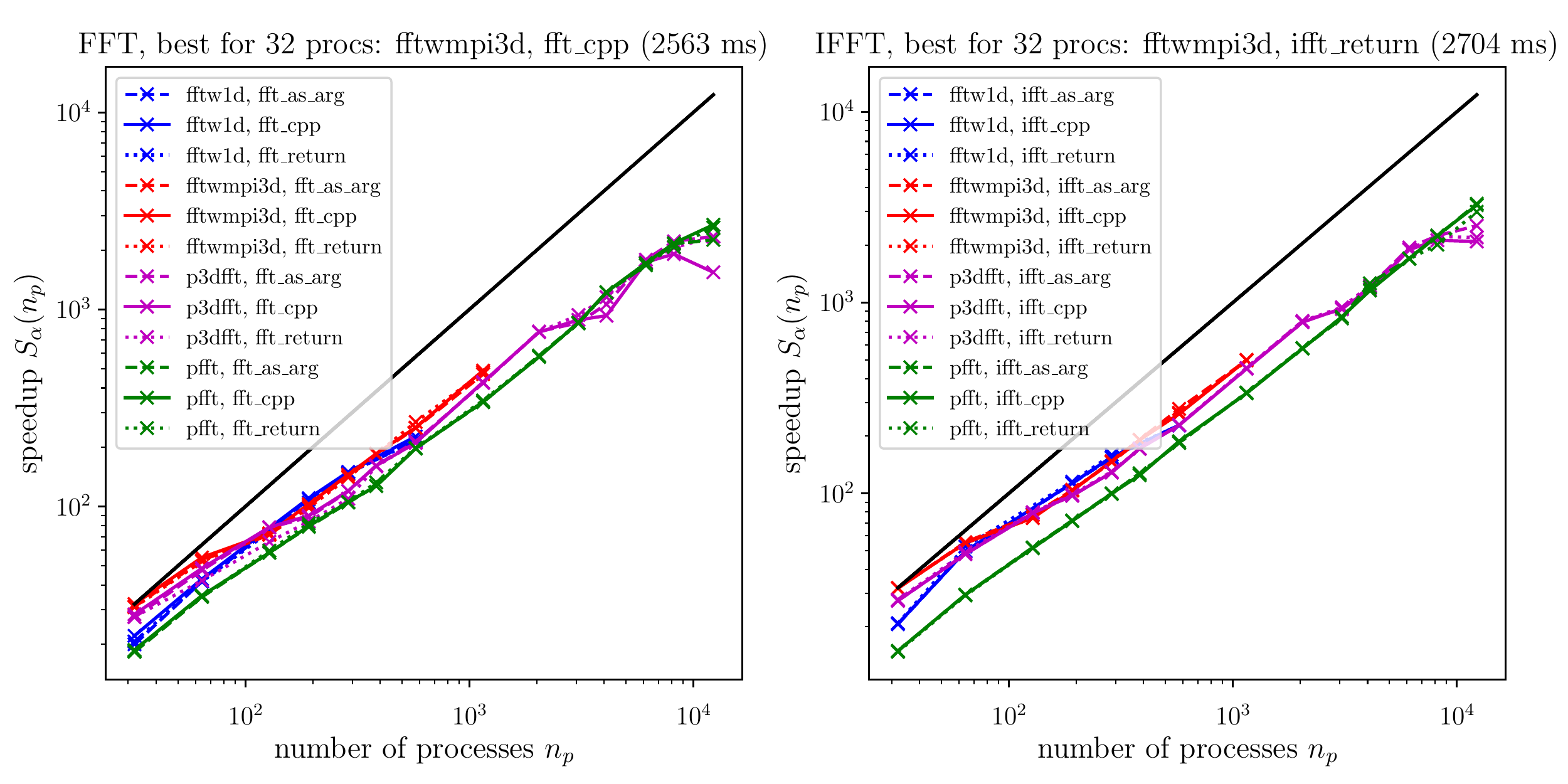}
\caption{Speedup computed from the median of the elapsed times for 3D fft
(1152$\times$1152$\times$1152, left: fft and right: ifft) on Beskow.}
\label{fig:beskow1152x1152x1152}
\end{figure}

The strong scaling results of the cubical array on Beskow are displayed on
Fig.~\ref{fig:beskow1152x1152x1152}, wherein we restrict to inter-node
computation.  We observe that the fastest method for low number of processes is
again, \codeinline{fftwmpi3d}. When high number of processes (above 1000)
are utilized, initially \codeinline{p3dfft} is the faster methods as before,
but with 3000 and above processes, \codeinline{pfft} is comparable in speed and
sometimes faster.

\paragraph{Benchmarks on a LEGI cluster}

Let us also analyse how \fluidpack{fft} scales on a computing cluster
maintained at an institutional level, named Cluster8 at \href{%
http://www.legi.grenoble-inp.fr}{LEGI}, Grenoble. This cluster functions using
Intel Xeon CPU E5-2650 v3 (2.3 GHz) with 20 cores per node and \fluidpack{fft}
was installed using a toolchain which comprises of gcc 4.9.2, Python 3.6.4 and
OpenMPI 1.6.5 as key software components.

\begin{figure}[htp!]
\centering
\includegraphics[width=\linewidth]{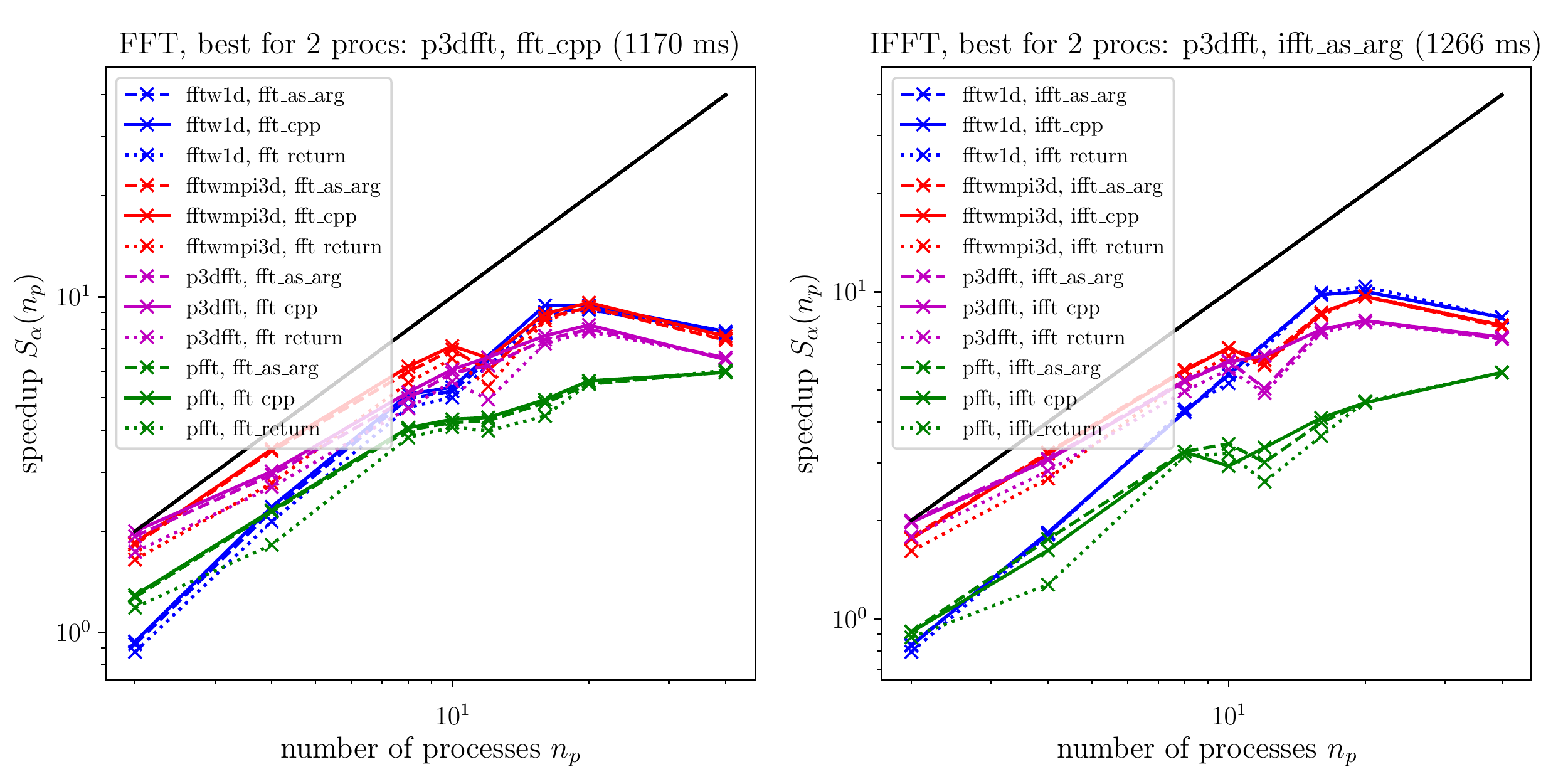}
\caption{Speedup computed from the median of the elapsed times for 3D fft
(320$\times$640$\times$640) at LEGI on cluster8.}
\label{fig:cluster8:320x640x640}
\end{figure}

In Fig.~\ref{fig:cluster8:320x640x640} we observe that the strong scaling for an
array shape of $320\times640\times640$ is not far from the ideal linear trend. The
fastest library is \codeinline{fftwmpi3d} for this case.  As expected from FFT
algorithms, there is a slight drop in speedup when the array size is not exactly
divisible by the number of processes, i.e.\ with 12 processes. The speedup
declines rapidly when more than one node is employed (above 20 processes). This
effect can be attributed to the latency introduced by inter-node communications, a
hardware limitation of this cluster (10 Gb/s).

\begin{figure}[htp!]
\centering
\includegraphics[width=\linewidth]{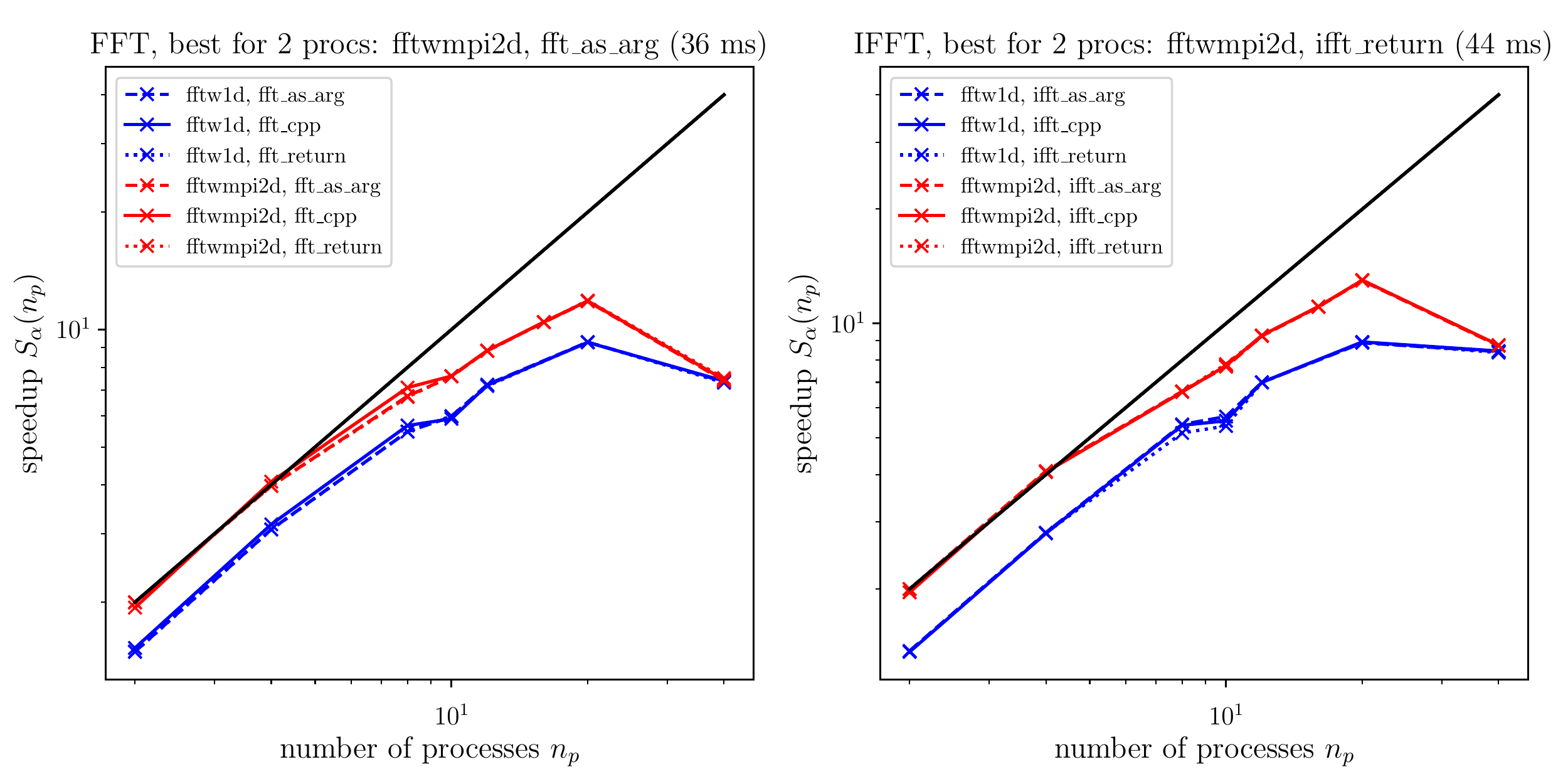}
\caption{Speedup computed from the median of the elapsed times for 2D fft
(2160$\times$2160) at LEGI on cluster8.}
\label{fig:cluster8:2160x2160}
\end{figure}

We have also analysed the performance of 2D MPI enabled FFT classes on the same
machine using an array shaped $2160\times2160$ in
Fig.~\ref{fig:cluster8:2160x2160}. The fastest library is
\codeinline{fftwmpi2d}. Both libraries display near-linear scaling, except when
more than one node is used and the performance tapers off.

As a conclusive remark on scalability, a general rule of thumb should be to use
1D domain decomposition when only very few processors are employed. For massive
parallelization, 2D decomposition is required to achieve good speedup without
being limited by the number of processors at disposal. We have thus shown that
overall performance of the libraries implemented in \fluidpack{fft} are quite
good, and there is no noticeable drop in speedup when the Python API is used.
This benchmark analysis also shows that the fastest FFT implementation depends
on the size of the arrays and on the hardware.
Therefore, an application build upon \fluidpack{fft} can be efficient for
different sizes and machines.

\subsubsection*{Microbenchmark of critical ``operator'' functions}

As already mentioned, we use \pack{Pythran} \citep{guelton2018pythran} to
compile some critical ``operator'' functions.  In this subsection, we present a
microbenchmark for one simple task used in pseudo-spectral codes: projecting a
velocity field on a non-divergent velocity field.  It is performed in spectral
space, where it can simply be written as
\begin{minted}[fontsize=\footnotesize]{python}
# pythran export proj_outplace(
#     complex128[][][], complex128[][][], complex128[][][],
#     float64[][][], float64[][][], float64[][][], float64[][][])

def proj_outplace(vx, vy, vz, kx, ky, kz, inv_k_square_nozero):
    tmp = (kx * vx + ky * vy + kz * vz) * inv_k_square_nozero
    return vx - kx * tmp, vy - ky * tmp, vz - kz * tmp
\end{minted}
Note that, this implementation is ``outplace'', meaning that the result is
returned by the function and that the input velocity field (\codeinline{vx, vy,
vz}) is unmodified.
The comment above the function definition is a \pack{Pythran} annotation, which
serves as a type-hint for the variables used within the functions --- all
arguments being \pack{Numpy} arrays in this case.
\pack{Pythran} needs such annotation to be able to compile this code into
efficient machine instructions \emph{via} a C++ code.
Without \pack{Pythran} the annotation has no effect, and
of course, the function defaults to using Python with \pack{Numpy} to execute.

The array notation is well adapted and less verbose to express this simple
vector calculus.
Since explicit loops with indexing is not required, the computation with Python
and \pack{Numpy} is not extremely slow. Despite this being quite a favourable
case for \pack{Numpy}, the computation with \pack{Numpy} is not optimized
because, internally, it involves many loops (one per arithmetic operator) and
creation of temporary arrays.

\begin{figure}[htp]
\centering
\includegraphics[width=\linewidth]{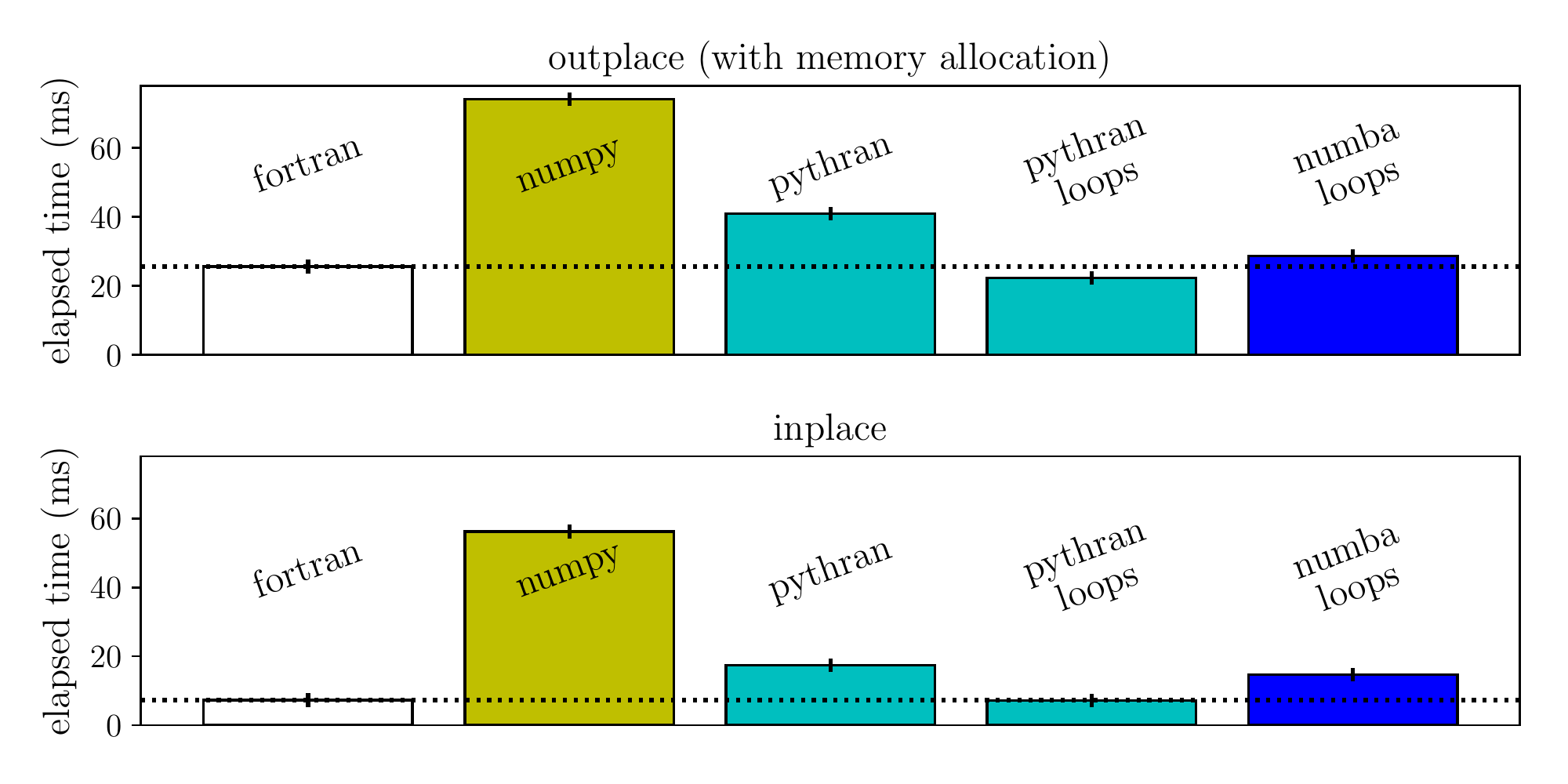}
\caption{Elapsed time (smaller is better) for the projection function for
different implementations and tools.  The shape of the arrays is
$(128,\ 128,\ 65)$. The dotted lines indicate the times for Fortran for better
comparison.}
\label{fig:microbench}
\end{figure}

In the top axis of Fig.~\ref{fig:microbench}, we compare the elapsed times for
different implementations of this function.
For this outplace version, we used three different codes:
\begin{enumerate}
\item a Fortran code (not shown\footnote{The codes and a Makefile used for this
benchmark study are available in \href{%
https://bitbucket.org/fluiddyn/fluiddyn_paper/src/default/fluidfft/microbench/}{%
the repository of the article}.}) written with three nested explicit loops (one
per dimension). Note that as in the Python version we also allocate the memory
where the result is stored.
\item the simplest Python version shown above.
\item a Python version with three nested explicit loops:
\begin{minted}[fontsize=\footnotesize]{python}
# pythran export proj_outplace_loop(
#     complex128[][][], complex128[][][], complex128[][][],
#     float64[][][], float64[][][], float64[][][], float64[][][])

def proj_outplace_loop(vx, vy, vz, kx, ky, kz, inv_k_square_nozero):

    rx = np.empty_like(vx)
    ry = np.empty_like(vx)
    rz = np.empty_like(vx)

    n0, n1, n2 = kx.shape

    for i0 in range(n0):
        for i1 in range(n1):
            for i2 in range(n2):
                tmp = (kx[i0, i1, i2] * vx[i0, i1, i2]
                       + ky[i0, i1, i2] * vy[i0, i1, i2]
                       + kz[i0, i1, i2] * vz[i0, i1, i2]
                ) * inv_k_square_nozero[i0, i1, i2]

                rx[i0, i1, i2] = vx[i0, i1, i2] - kx[i0, i1, i2] * tmp
                ry[i0, i1, i2] = vz[i0, i1, i2] - kx[i0, i1, i2] * tmp
                rz[i0, i1, i2] = vy[i0, i1, i2] - kx[i0, i1, i2] * tmp

    return rx, ry, rz
\end{minted}
\end{enumerate}
For the version without explicit loops, we present the elapsed time for two
cases: (i) simply using Python (yellow bar) and (ii) using the Pythranized
function (first cyan bar).
For the Python version with explicit loops, we only present the results for (i)
the Pythranized function (second cyan bar) and (ii) the result of \pack{Numba}
(blue bar).
We do not show the result for \pack{Numba} for the code without explicit loops
because it is slower than \pack{Numpy}. We have also omitted the result for
\pack{Numpy} for the code with explicit loops because it is very inefficient.
The timing is performed upon tuning the computer using the package
\href{https://pypi.org/project/perf/}{\pack{perf}}.

We see that \pack{Numpy} is approximately three time slower than the Fortran
implementation (which as already mentioned contains the memory allocation).
Just using \pack{Pythran} without changing the code (first cyan bar), we save
nearly 50\% of the execution time but we are still significantly slower than
the Fortran implementation.
We reach the Fortran performance (even slightly faster) only by using
\pack{Pythran} with the code with explicit loops.
With this code, \pack{Numba} is nearly as fast (but still slower) without
requiring any type annotation.

Note that the exact performance differences depend on the hardware, the software
versions\footnote{Here, we use Python~3.6.4 (packaged by conda-forge),
\pack{Numpy}~1.13.3, \pack{Pythran}~0.8.5, \pack{Numba}~0.38, gfortran~6.3 and
clang~6.0.}, the compilers and the compilation options.
We use \codeinline{gfortran -O3 -march=native} for Fortran and
\codeinline{clang++ -O3 -march=native} for \pack{Pythran}\footnote{The results
with \codeinline{g++ -O3 -march=native} are very similar but tend to be slightly
slower.}.

Since allocating memory is expensive and we do not need the non-projected
velocity field after the call of the function, an evident optimization is to
put the output in the input arrays.  Such an ``in-place'' version can be written
with \pack{Numpy} as:
\begin{minted}[fontsize=\footnotesize]{python}
# pythran export proj_inplace(
#     complex128[][][], complex128[][][], complex128[][][],
#     float64[][][], float64[][][], float64[][][], float64[][][])

def proj_inplace(vx, vy, vz, kx, ky, kz, inv_k_square_nozero):
    tmp = (kx * vx + ky * vy + kz * vz) * inv_k_square_nozero
    vx -= kx * tmp
    vy -= ky * tmp
    vz -= kz * tmp
\end{minted}

As in the first version, we have included the \pack{Pythran} annotation.
We also consider an ``in-place'' version with explicit loops:
\begin{minted}[fontsize=\footnotesize]{python}
# pythran export proj_inplace_loop(
#     complex128[][][], complex128[][][], complex128[][][],
#     float64[][][], float64[][][], float64[][][], float64[][][])

def proj_inplace_loop(vx, vy, vz, kx, ky, kz, inv_k_square_nozero):

    n0, n1, n2 = kx.shape

    for i0 in range(n0):
        for i1 in range(n1):
            for i2 in range(n2):
                tmp = (kx[i0, i1, i2] * vx[i0, i1, i2]
                       + ky[i0, i1, i2] * vy[i0, i1, i2]
                       + kz[i0, i1, i2] * vz[i0, i1, i2]
                ) * inv_k_square_nozero[i0, i1, i2]

                vx[i0, i1, i2] -= kx[i0, i1, i2] * tmp
                vy[i0, i1, i2] -= ky[i0, i1, i2] * tmp
                vz[i0, i1, i2] -= kz[i0, i1, i2] * tmp

\end{minted}
Note that this code is much longer and clearly less readable than the version
without explicit loops.  This is however the version which is used in
\pack{fluidfft} since it leads to faster execution.

The elapsed time for these inplace versions and for an equivalent Fortran
implementation are displayed in the bottom axis of Fig.~\ref{fig:microbench}.
The ranking is the same as for the outplace versions and \pack{Pythran} is also
the faster solution.
However, \pack{Numpy} is even more slower (7.8 times slower than \pack{Pythran}
with the explicit loops) than for the outplace versions.

From this short and simple microbenchmark, we can infer four main points:
\begin{itemize}
\item Memory allocation takes time!  In Python, memory management is automatic
and we tend to forget it.  An important rule to write efficient code is to
reuse the buffers already allocated as much as possible.

\item Even for this very simple case quite favorable for \pack{Numpy} (no indexing
or slicing), \pack{Numpy} is three to eight time slower than the Fortran
implementations. As long as the execution time is small or that the
function represents a small part of the total execution time, this is not an
issue. However, in other cases, Python-\pack{Numpy} users need to consider
other solutions.

\item \pack{Pythran} is able to speedup the \pack{Numpy} code without explicit
loops and is as fast as Fortran (even slightly faster in our case) for the
codes with explicit loops.

\item \pack{Numba} is unable to speedup the \pack{Numpy} code.
It gives very interesting performance for the version with explicit loops
without any type annotation but the result is significantly slower than with
\pack{Pythran} and Fortran.
\end{itemize}

\section*{Quality control}


The package \fluidpack{fft} currently supplies unit tests covering 93\% of its
code.  These unit tests are run regularly through continuous integration on Travis
CI with the most recent releases of \fluidpack{fft}'s dependencies and on
Bitbucket Pipelines inside a static
\href{https://hub.docker.com/u/fluiddyn}{Docker container}.  The tests are run
using standard Python interpreter with all supported versions.

For \fluidpack{fft}, the code coverage results are displayed at
\href{https://codecov.io/bb/fluiddyn/fluidfft}{Codecov}.  Using third-party
packages \pack{coverage} and \pack{tox}, it is straightforward to bootstrap the
installation with dependencies, test with multiple Python versions and combine the
code coverage report, ready for upload. It is also possible to run similar
isolated tests using \pack{tox} or coverage analysis using \pack{coverage} in a
local machine.  Up-to-date build status and coverage status are displayed on the
landing page of the Bitbucket repository.  Instructions on how to run unit tests,
coverage and lint tests are included in the documentation.

We also try to follow a consistent code style as recommended by PEP (Python
enhancement proposals) 8 and 257. This is also inspected using lint checkers such
as \codeinline{flake8} and \codeinline{pylint} among the developers.  The Python
code is regularly cleaned up using the code formatter \codeinline{black}.

\section*{(2) Availability}
\vspace{0.5cm}
\section*{Operating system}


Windows and any POSIX based OS, such as GNU/Linux and macOS.

\section*{Programming language}


Python 2.7, 3.5 or above.
Note that while Cython and Pythran both use the C API of CPython, \fluidpack{fft}
has been successfully tested on PyPy 6.0.
A C++11 supporting compiler, while not mandatory for the C++ API or Cython
extensions of \fluidpack{fft}, is recommended to able to use Pythran extensions.

\section*{Dependencies}


\begin{itemize}
\item {\bf Minimum:} \fluidpack{dyn}, \pack{Numpy}, \pack{Cython}, and
  \pack{mako}\ or \pack{Jinja2}; \libpack{FFTW} library.
\item {\bf Optional:} \pack{mpi4py} and \pack{Pythran}; \libpack{P3DFFT},
  \libpack{PFFT} and \libpack{cuFFT} libraries.
\end{itemize}

\section*{List of contributors}


\begin{itemize}
\item Pierre Augier (LEGI): creator of the FluidDyn project and of
\fluidpack{fft}.
\item Cyrille Bonamy (LEGI): C++ code and some methods in the operator classes.
\item Ashwin Vishnu Mohanan (KTH): command lines utilities, benchmarks, unit
  tests, continuous integration, and bug fixes.
\end{itemize}

\section*{Software location:}


\begin{description}[noitemsep,topsep=0pt]
\item[Name:] PyPI
\item[Persistent identifier:] https://pypi.org/project/fluidfft
\item[Licence:] CeCILL, a free software license adapted to both international
and French legal matters, in the spirit of and retaining compatibility with the
GNU General Public License (GPL).
\item[Publisher:] Pierre Augier
\item[Version published:] 0.2.4
\item[Date published:] 02/07/2018
\end{description}

{\bf Code repository}

\begin{description}[noitemsep,topsep=0pt]
\item[Name:] Bitbucket
\item[Persistent identifier:] https://bitbucket.org/fluiddyn/fluidfft
\item[Licence:] CeCILL
\item[Date published:] 2017
\end{description}

{\bf Emulation environment}

\begin{description}[noitemsep,topsep=0pt]
\item[Name:] Docker
\item[Persistent identifier:] https://hub.docker.com/r/fluiddyn/python3-stable
\item[Licence:] CeCILL-B, a BSD compatible French licence.
\item[Date published:] 02/10/2017
\end{description}

\section*{Language}


English

\section*{(3) Reuse potential}


\fluidpack{fft} is used by the Computational Fluid Mechanics framework
\fluidpack{sim} \citep{fluidsim}. It could be used by any C++ or Python project
where real-to-complex 2D or 3D FFTs are performed.

There is no formal support mechanism. However, bug reports can be submitted at
the \href{https://bitbucket.org/fluiddyn/fluidsim/issues}{Issues page on
Bitbucket}. Discussions and questions can be aired on instant messaging
channels in Riot (or equivalent with Matrix protocol)\footnote{
\url{%
  https://matrix.to/\#/\#fluiddyn-users:matrix.org}}
or via IRC protocol on Freenode at \codeinline{\#fluiddyn-users}. Discussions
can also be exchanged via the official mailing list\footnote{
\url{https://www.freelists.org/list/fluiddyn}}.

\section*{Acknowledgements}


Ashwin Vishnu Mohanan could not have been as involved in this project without the
kindness of Erik Lindborg.
We are grateful to Bitbucket for providing us with a high quality forge
compatible with Mercurial, free of cost.

\section*{Funding statement}


This project has indirectly benefited from funding from the foundation Simone et
Cino Del Duca de l'Institut de France, the European Research Council (ERC)
under the European Union's Horizon 2020 research and innovation program (grant
agreement No 647018-WATU and Euhit consortium) and the Swedish Research Council
(Vetenskapsr{\aa}det): 2013--5191.
We have also been able to use supercomputers of CIMENT/GRICAD, CINES/GENCI (Grant
2018-A0040107567) and the Swedish National Infrastructure for Computing (SNIC).

\section*{Competing interests}


The authors declare that they have no competing interests.



\bibliographystyle{agsm}
\bibliography{bib}

\rule{\textwidth}{1pt}

{\bf Copyright Notice} \\
Authors who publish with this journal agree to the following terms: \\

Authors retain copyright and grant the journal right of first publication with
the work simultaneously licensed under a
\href{http://creativecommons.org/licenses/by/3.0/}{Creative Commons Attribution
License} that allows others to share the work with an acknowledgement of the
work's authorship and initial publication in this journal.

Authors are able to enter into separate, additional contractual arrangements
for the non-exclusive distribution of the journal's published version of the
work (e.g., post it to an institutional repository or publish it in a book),
with an acknowledgement of its initial publication in this journal.

By submitting this paper you agree to the terms of this Copyright Notice, which
will apply to this submission if and when it is published by this journal.

\end{document}